\documentclass[letterpaper, 10 pt, conference]{ieeeconf}  % Comment this line out if you need a4paper

\IEEEoverridecommandlockouts                              % This command is only needed if 
\usepackage{amsmath} % assumes amsmath package installed
\usepackage[T1]{fontenc}
\usepackage{graphicx}
\usepackage{amssymb}
\usepackage{caption}
\usepackage{subcaption}
\usepackage{float}

\title{\LARGE \bf
Adaptive Wall-Following Control for Unmanned Ground Vehicles Using Spiking Neural Networks
}

\author{Hengye Yang, Yanxiao Chen, Zexuan Fan, Lin Shao and Tao Sun%
\thanks{Robotics Algorithm Team, Microwave $\&$ Cleaning Appliance Division, Midea Group
       {\tt\small hengyeyang@gmail.com; \{chenyx190, zexuan.fan, lin.shao, tsun\}@midea.com}}%
}

\begin{document}

\maketitle
\thispagestyle{empty}
\pagestyle{empty}

%%%%%%%%%%%%%%%%%%%%%%%%%%%%%%%%%%%%%%%%%%%%%%%%%%%%%%%%%%%%%%%%%%%%%%%%%%%%%%%%
\begin{abstract}
Unmanned ground vehicles operating in complex environments must adaptively adjust to modeling uncertainties and external disturbances to perform tasks such as wall following and obstacle avoidance. This paper introduces an adaptive control approach based on spiking neural networks for wall fitting and tracking, which learns and adapts to unforeseen disturbances. We propose real-time wall-fitting algorithms to model unknown wall shapes and generate corresponding trajectories for the vehicle to follow. A discretized linear quadratic regulator is developed to provide a baseline control signal based on an ideal vehicle model. Point matching algorithms then identify the nearest matching point on the trajectory to generate feedforward control inputs. Finally, an adaptive spiking neural network controller, which adjusts its connection weights online based on error signals, is integrated with the aforementioned control algorithms. Numerical simulations demonstrate that this adaptive control framework outperforms the traditional linear quadratic regulator in tracking complex trajectories and following irregular walls, even in the presence of partial actuator failures and state estimation errors.

\end{abstract}

%%%%%%%%%%%%%%%%%%%%%%%%%%%%%%%%%%%%%%%%%%%%%%%%%%%%%%%%%%%%%%%%%%%%%%%%%%%%%%%%
\section{Introduction}
\label{sec:Intro}
Unmanned ground vehicles (UGVs) are extensively utilized in applications such as military surveillance, agricultural irrigation, and floor cleaning~\cite{CaoUGVMPPlanning24}. As operational environments become more complex, UGVs must effectively explore unknown areas and adapt to uncertainties. On-board wall modeling and following are essential for efficient exploration, as accurate boundary modeling enhances subsequent motion planning. Fuzzy logic control, which uses human-defined rules, is a common method for complex wall following~\cite{LeeFuzzyCtrl17}. For example, \cite{LiFuzzyCtrl15} describes a behavior-based fuzzy controller that enables a mobile robot to navigate walls with both convex and concave corners. 
% Additionally, a reinforcement-learning approach in~\cite{JuangRLFuzzyCtrl09} selects actions based on ant pheromone trails and Q-values, updated through reinforcement signals. 
However, fuzzy logic controllers have a significant limitation: their design and tuning depend heavily on human expertise, making them less adaptable to unforeseen circumstances with limited fuzzy data sets.

Modeling wall shapes using high-quality sensor data and generating corresponding curve trajectories for tracking has emerged as a prominent approach in wall-following control. The effectiveness of wall-following in complex environments significantly depends on the precision of trajectory tracking amidst modeling uncertainties and environmental disturbances~\cite{YangPhDThesis23}. Beyond traditional PID controllers~\cite{XuPID22}, an adaptive model-based trajectory tracking controller with a dynamic parameter-updating law was developed in~\cite{MartinsAdaDynCtrl08}. Additionally, Model Predictive Control (MPC) integrates both the vehicle model and environmental constraints to derive an optimal control strategy for trajectory tracking~\cite{NascimentoMPCSurvey18}. For instance, \cite{DingMPCVelCtrl22} introduces a steer-based MPC combined with a steering fuzzy selector to optimize wheel velocity and achieve rapid tracking convergence. However, these control methods often depend on precise kinematic and dynamic vehicle models or require prior identification of disturbance models, thereby limiting their real-time effectiveness when vehicle models become inaccurate due to drastic environmental changes, such as rough and uneven terrain.

To address modeling uncertainties and external disturbances, control systems must be adaptively reconfigured by either modifying the control loop structure or adjusting system parameters~\cite{StengelIntelligentFlightCtrl93}. Artificial neural networks (ANNs) are particularly advantageous for reconfigurable control design due to their ability to dynamically adjust connection weights in response to discrepancies between desired and actual system responses~\cite{FerrariAdaptiveNNLearning02}. 
For example, \cite{FerrariAdaptiveCriticCtrl04} presents a neural-network-based control framework that incorporates an offline gain-scheduled controller and an online dual heuristic adaptive critic controller, effectively handling unforeseen conditions such as physical parameter variations and modeling uncertainties. 
For instance, \cite{HassanNNAdaCtrl22} introduces a hybrid control approach combining an ANN-based kinematic controller with a model reference adaptive controller, outperforming PID controllers in tracking accuracy and convergence speed. However, the ANNs that provide gains for the kinematic controller require meticulous offline training for a specific plant, rendering them unsuitable for rapidly changing environments and scenarios with limited computational resources.

Spiking neural networks (SNNs) emulate the functioning of biological brains by transmitting information through discrete-time spikes when a neuron's membrane potential exceeds a threshold~\cite{YamazakiSNNReview22, YangRelativeSpikeTimingKernel22}. This energy-efficient mechanism makes SNN controllers ideal for modern small-scale UGVs equipped with neuromorphic chips, potentially replacing traditional ANN-based controllers~\cite{RathiSNNHardware23}. Therefore, this paper presents an adaptive SNN-based wall-following control framework that integrates real-time wall-fitting, a baseline linear quadratic regulator (LQR) for trajectory tracking, a feedforward point-matching module to accelerate convergence, and an SNN-based feedback controller capable of learning and adapting to modeling uncertainties and external disturbances. The novelty of this work lies in its ability to operate without prior identification of the disturbance model, making it suitable for various types of uncertainties. Additionally, both the feedforward point matching and SNN adaptation enhance real-time convergence speed. This adaptive SNN controller achieves accurate trajectory tracking and wall following despite significant uncertainties, including partial actuator failure and state estimation errors.

This paper is organized as follows. Section~\ref{sec:ProbForm} formulates the wall-following and trajectory tracking problem. Section~\ref{sec:Method} introduces the wall-fitting approach and adaptive control design. Section~\ref{sec:Results} presents the numerical simulation results. Finally, Section~\ref{sec:Conclusion} provides the conclusion.

\section{Problem Formulation}
\label{sec:ProbForm}
Traditional UGVs operating in complex and dynamic environments are highly sensitive to modeling uncertainties and external disturbances. Tasks such as irregular or discontinuous wall following, obstacle avoidance, and complex trajectory tracking require onboard controllers to adaptively adjust their motion based on onboard sensing. This paper addresses the problem of developing a robust control law for UGVs that can adapt to common uncertainties, including partial actuator failures and estimation errors. Typically, the UGV's nonlinear dynamic model can be written as
\begin{equation}
\label{eq:OriginVehicleDyn}
    \dot{\mathbf{x}}(t)=\mathbf{f}[\mathbf{x}(t),\mathbf{u}(t)],\;
    \mathbf{x}(t_0)=\mathbf{x}_0
\end{equation}
where $\mathbf{x}$ denotes the vehicle state, and $\mathbf{u}$ represents the control input. Given the dynamic constraints in~\eqref{eq:OriginVehicleDyn}, onboard state measurements $\hat{\mathbf{x}}$, and desired state $\mathbf{x}_r$, we aim to determine a bounded control history $\mathbf{u}(t)$ that enables the vehicle to track a desired trajectory specified by wall-fitting algorithms. The tracking error $\mathbf{e}(t)=\|\hat{\mathbf{x}}(t)-\mathbf{x}_r(t)\|$, resulting from modeling uncertainties, remains within acceptable limits and asymptotically approaches zero. The desired optimal control law $\mathbf{u}$ needs to be robust to significant uncertainties and minimize the cost function below:
\begin{equation}
\label{eq:GeneralCostFunc}
    J = \phi[\mathbf{x}(t_f)] + 
    \int_{t_0}^{t_f} L[\mathbf{x}(t),\mathbf{u}(t)] dt
\end{equation}
where $\phi$ is the terminal cost, and the Lagrangian $L$ may include quadratic state deviation or control cost, depending on the specific control objectives, as elaborated in the subsequent section on discretized linear quadratic regulation.

\section{Methodology}
\label{sec:Method}
This section presents the wall-fitting and adaptive SNN-based trajectory following approaches to address the wall-following problem. In Section \ref{ssec:BsplineFitting}, a B-spline wall-fitting method is developed to generate a reference trajectory for the UGV. Section \ref{ssec:LQR} derives a baseline LQR control solution $\mathbf{u}_l$ for trajectory tracking. In Section \ref{ssec:PtMatching}, a matching point finder (MPF) is developed to generate feedforward control signals $\mathbf{u}_f$. Finally, Section \ref{ssec:AdaSNN} details the derivation of an adaptive SNN control solution $\mathbf{u}_a$, which is combined with LQR state feedback and feedforward control signals to achieve trajectory tracking objectives despite modeling uncertainties and external disturbances. The overall control architecture is illustrated in Fig.~\ref{fig:BlockDiagram}.

\subsection{B-spline Wall Fitting}
\label{ssec:BsplineFitting}
Initially, onboard LiDAR data is collected as a point cloud, $\{P_0, P_1, \cdots, P_N\}$, comprising $N+1$ points reflected from the nearby wall surface. We then use a $k$-degree B-spline to model the shape of the wall surface as follows:
\begin{equation}
\label{eq:B-spline function}
    P(t)=\sum_{i=0}^N N_{i,k}(t)P_i
\end{equation}
where each parameter $t\in[0,1]$ corresponds to the position $P$ of a point on the B-spline \cite{UnserBsplineTheory93}. For a quasi-uniform B-spline, the knot vector can be represented as
\begin{equation}
\label{eq:knot vector}
\begin{split}
    T &= [t_0,\cdots,t_k,t_{k+1},\cdots,t_n,t_{n+1},\cdots,t_{n+k+1}]\\
      &= [0,\cdots,0,\tfrac{1}{n-k+1},\cdots,\tfrac{n-k}{n-k+1},1,\cdots,1]
\end{split}
\end{equation}
where $k$ duplicate knots are added before $t_k=0$ and after $t_{n+1}=1$ to ensure tangency at both ends, and the $n-k+2$ interior knots are uniformly spaced with a knot span of $\Delta t=\frac{1}{n-k+1}$. The base function $N_{i,k}$ can then be calculated as
\begin{subequations}
\begin{equation}
\label{eq:base function}
    N_{i,k}(t) = \frac{t-t_i}{t_{i+k}-t_i} N_{i,k-1}(t) + \frac{t_{i+k+1}-t}{t_{i+k+1}-t_{i+1}} N_{i+1,k-1}(t)
\end{equation}
\begin{equation}
\label{eq:base function initial}
    N_{i,0}(t) =
    \begin{cases}
    1 & \quad t\in[t_i,t_{i+1})\\
    0 & \quad \text{otherwise}
    \end{cases}
\end{equation}
\end{subequations}
We choose B-spline curve fitting over Bézier or classical polynomial curve fitting approaches because the control points of a B-spline curve influence only a local region, preserving the overall shape.

\subsection{Linear Quadratic Regulation}
\label{ssec:LQR}
Based on the results of the B-spline wall fitting, a corresponding trajectory is generated for the vehicle to follow while maintaining a specified distance from the wall. We assume that the trajectory can be represented by the following parametric equations:
\begin{equation}
\label{eq:para eqn}
    \begin{cases}
    x=\phi(t) \\
    y=\psi(t)
    \end{cases}
\end{equation}
In this paper, we consider the trajectory tracking problem of a UGV, which can be modeled as a differential drive model. The vehicle's linear velocity $v$ and angular velocity $\omega$ are determined by the rotational velocities of its right and left wheels, $v_R$ and $v_L$, respectively:
\begin{equation}
    v=\frac{v_R+v_L}{2},\;\omega=\frac{v_L-v_R}{L}
\end{equation}
where $L$ denotes the wheel distance. The dynamic model of the vehicle can be given by
\begin{equation}
\label{eq:dyn model}
\begin{bmatrix} \dot{x}(t) \\ \dot{y}(t) \\ \dot{\theta}(t) \end{bmatrix}=
\begin{bmatrix}
    \cos{\theta(t)} & 0 \\
    \sin{\theta(t)} & 0 \\
    0 & 1
\end{bmatrix}
\begin{bmatrix}
    v(t) \\ w(t)
\end{bmatrix}
\end{equation}
where $\mathbf{x}=[x\;y\;\theta]^T$ denotes the state vector, and $\mathbf{u}=[v\;w]^T$ represents the control input. The method to find the closest reference trajectory point and compute a feedforward portion of the control input in real time will be illustrated in the following subsection. Given the reference trajectory point $\mathbf{x}_r=[x_r\;y_r\;\theta_r]^T$ and control input $\mathbf{u}_r=[v_r\;\omega_r]^T$, the nonlinear dynamic model can be linearized around the reference trajectory as
\begin{equation}
\label{eq:ss eqn}
    \dot{\tilde{\mathbf{x}}}(t)=\mathbf{A}\tilde{\mathbf{x}}(t)
        +\mathbf{B}\tilde{\mathbf{u}}(t)
\end{equation}
where $\tilde{\mathbf{x}}=\mathbf{x}-\mathbf{x}_r$, $\tilde{\mathbf{u}}=\mathbf{u}-\mathbf{u}_r$, and the state-space matrices are given by
\begin{equation}
\label{eq:ss matrices}
    \mathbf{A}=\begin{bmatrix}
        0 & 0 & -v_r\sin{\theta_r} \\
        0 & 0 & v_r\cos{\theta_r} \\
        0 & 0 & 0
    \end{bmatrix};
    \mathbf{B}=\begin{bmatrix}
        \cos{\theta_r} & -v_r\sin{\theta_r}\delta t \\
        \sin{\theta_r} & v_r\cos{\theta_r}\delta t \\
        0 & 1
    \end{bmatrix}
\end{equation}
where $\delta t$ is the sampling period. Discretizing~\eqref{eq:ss eqn} using the forward Euler method, the discretized vehicle dynamic model can be expressed as
\begin{equation}
    \tilde{\mathbf{x}}_{k+1}=
    (\delta t\mathbf{A}+\mathbf{I})\tilde{\mathbf{x}}_k+
    (\delta t\mathbf{B})\tilde{\mathbf{u}}_k
    =\mathbf{A}_k\tilde{\mathbf{x}}_k+\mathbf{B}_k\tilde{\mathbf{u}}_k
\end{equation}
To find the optimal trajectory tracking control history, we need to minimize the quadratic cost function:
\begin{equation}
\label{eq:cost func}
    J=\frac{1}{2}\sum\limits_{k=1}^{N-1}
    (\tilde{\mathbf{x}}_k^T\mathbf{Q}\tilde{\mathbf{x}}_k
    +\tilde{\mathbf{u}}_k^T\mathbf{R}\tilde{\mathbf{u}}_k)
    +\frac{1}{2}\tilde{\mathbf{x}}_N^T\mathbf{Q}_f\tilde{\mathbf{x}}_N
\end{equation}
where the first two terms quantify the state deviation from the reference point and the control effort during the process, respectively, and the third term represents the final state deviation~\cite{StengelOCEbook94}. The weighting matrices $\mathbf{Q}\in\mathbb{R}^{3\times3}$, $\mathbf{Q}_f\in\mathbb{R}^{3\times3}$, and $\mathbf{R}\in\mathbb{R}^{2\times2}$ are all positive semi-definite. As the horizon $N$ approaches infinity, the discretized LQR optimal state feedback control law can be derived as
\begin{equation}
    \mathbf{u}_l(t_k)\triangleq\tilde{\mathbf{u}}_k=-\mathbf{K}_{ss}\tilde{\mathbf{x}}_k
\end{equation}
The steady-state feedback gain matrix $\mathbf{K}_{ss}$ is given by
\begin{equation}
    \mathbf{K}_{ss}=(\mathbf{R}+\mathbf{B}_k^T\mathbf{P}_{ss}\mathbf{B}_k)^{-1}\mathbf{B}_k^T\mathbf{P}_{ss}\mathbf{A}_k
\end{equation}
and $\mathbf{P}_{ss}$ can be determined via iterative calculations of the Algebraic Riccati Equation (ARE) below until convergence is attained:
\begin{equation}
\begin{split}
    \mathbf{P}_{n+1}=-\mathbf{A}_k^T\mathbf{P}_n\mathbf{B}_k
    (\mathbf{R}+\mathbf{B}_k^T\mathbf{P}_n\mathbf{B}_k)^{-1}
    \mathbf{B}_k^T\mathbf{P}_n\mathbf{A}_k \\
    +\mathbf{Q}+\mathbf{A}_k^T\mathbf{P}_n\mathbf{A}_k
\end{split}
\end{equation}

\begin{figure}[ht!]
\centering
\includegraphics[width=0.48\textwidth]{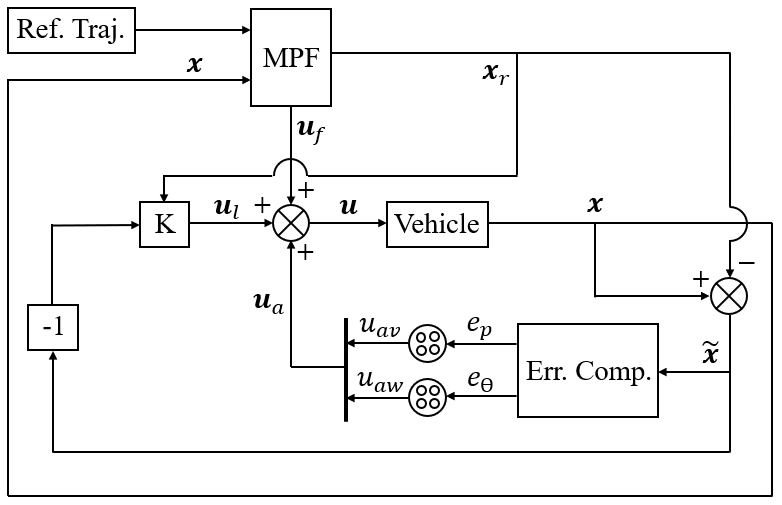}
\caption{Framework of the adaptive SNN control design.}
\label{fig:BlockDiagram}
\end{figure}

\subsection{Feedforward Point Matching}
\label{ssec:PtMatching}
To track the reference trajectory, the cost function defined in~\eqref{eq:cost func} minimizes the deviation of the state from the reference point $\mathbf{x}_r$. In MPF, the point index $m_r$ is determined in real time by identifying the closest point to the vehicle's current position $(x,y)$ within a user-defined range:
\begin{equation}
    m_r=\mathop{argmin}_m [(x-x_m)^2+(y-y_m)^2]
\end{equation}
Since the reference point $\mathbf{x}_r$ evolves over time, the state-space matrices in~\eqref{eq:ss matrices} are updated by MPF as the vehicle approaches the trajectory. To accelerate convergence, the curvature at each matching point on the parametric curve defined in~\eqref{eq:para eqn} is calculated as
\begin{equation}
    \kappa=\frac{|\phi'(t)\psi''(t)-\phi''(t)\psi'(t)|}
    {[\phi'^2(t)+\psi'^2(t)]^{\frac{3}{2}}}
\end{equation}
which is then provided to the vehicle as a feedforward control signal, $\mathbf{u}_f=[v_r\hspace{2mm}\alpha\kappa]^T$, where $v_r$ is a user-defined constant velocity, and $\alpha$ is a feedforward coefficient.

\subsection{Adaptive Control Design}
\label{ssec:AdaSNN}
\begin{figure}[ht!]
\centering
\includegraphics[width=0.26\textwidth]{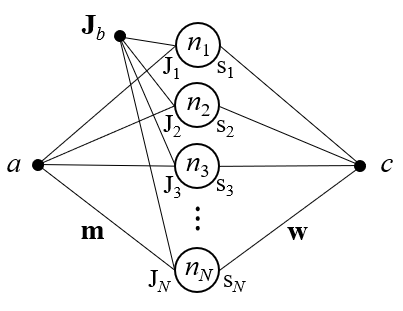}
\caption{Single-layer SNN architecture.}
\label{fig:SNN}
\end{figure}

LQR is designed to stabilize an ideal vehicle model that has been linearized around the desired trajectory. However, this approach may fail if the vehicle is initially positioned too far from the desired path or if the vehicle model lacks sufficient accuracy~\cite{YangPhDThesis23, StengelOCEbook94}. To address these challenges, we develop an additional adaptive SNN controller to compensate for unexpected disturbances. Fig.~\ref{fig:SNN} illustrates the architecture of a single-layer SNN. According to the Neural Engineering Framework (NEF)~\cite{EliasmithNeuralEngineering03}, the continuous-time input signal $a$ is first encoded into $N$ spiking neurons, and the input current $\mathbf{J}$ supplied to the neurons is defined by
\begin{equation}
    \mathbf{J}(t)=\mathbf{m}a(t)+\mathbf{J}_b
\end{equation}
where $\mathbf{m}$ represents the input connection (encoding) weights, and $\mathbf{J}_b$ is a fixed bias current. Then, the output signal $c$ is decoded from the post-synaptic current $\mathbf{s}$ of these neurons,
\begin{equation}
    c(t)=\mathbf{w}^T\mathbf{s}(t)=\mathbf{w}^TF[\mathbf{J}(t)]
    =\mathbf{w}^TF[\mathbf{m}a(t)+\mathbf{J}_b]
\end{equation}
where $\mathbf{w}$ is the output connection (decoding) weights, and $F(\cdot)$ is the nonlinear activation function. 

\begin{figure}[ht!]
\centering
\includegraphics[width=0.4\textwidth]{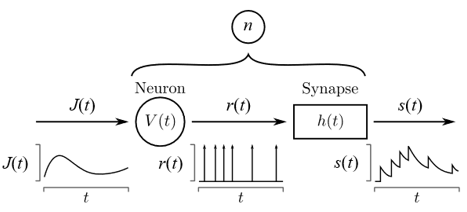}
\caption{Leaky Integrate-and-Fire (LIF) neuron model.}
\label{fig:LIFNeuronModel}
\end{figure}

Although the structure of the SNN presented above resembles that of a conventional ANN, the modeling of individual spiking neurons is significantly more complex. In this paper, we employ the Leaky Integrate-and-Fire (LIF) model as described in~\cite{GerstnerSpikingNeuronModels02} to represent the dynamics of spiking neurons. As illustrated in Fig.~\ref{fig:LIFNeuronModel}, the neuron's membrane voltage $V$ depends on the input current $J$,
\begin{equation}
    \dot{V}(t)=-\frac{1}{\tau_d}[V(t)-RJ(t)]
\end{equation}
where $\tau_d$ represents the decaying time constant, and $R$ denotes the passive membrane resistance. Spikes are emitted when the membrane voltage exceeds a threshold, and the resulting spike trains $r$ can be represented as a sum of Dirac delta functions $\delta$,
\begin{equation}
    r(t)=F[J(t)]=\sum\limits_k\delta(t-t_k)
\end{equation}
where $k$ denotes the spike index. Then, synapses filter the spike trains via an exponential decaying function $h(t)=\frac{1}{\tau_p}e^{-t/\tau_p}$, and generate post-synaptic currents,
\begin{equation}
    s(t)=r(t)\ast h(t)=\sum\limits_kh(t-t_k)
\end{equation}
where $\tau_p$ is the post-synaptic time constant.

The adaptive control signal $\mathbf{u}_a$ consists of two terms: the adaptive linear velocity input $u_{av}$ and the angular velocity input $u_{aw}$,
\begin{equation}
    \mathbf{u}_a=[u_{av}\;u_{aw}]^T
\end{equation}
Each element of the adaptive control signal is computed from a single network of $100$ spiking neurons,
\begin{subequations}
\begin{equation}
    u_{av}(t)=\mathbf{w}_v^T\mathbf{s}_v(t)
\end{equation}
\begin{equation}
    u_{aw}(t)=\mathbf{w}_w^T\mathbf{s}_w(t)
\end{equation}
\end{subequations}
Through the error computation module in Fig.~\ref{fig:BlockDiagram}, the position error signal $e_p$ is dependent on the position deviation from the closest reference trajectory point:
\begin{equation}
    e_p=\sqrt{(x-x_r)^2+(y-y_r)^2}
\end{equation}
Subsequently, using the Prescribed Error Sensitivity (PES) learning rule~\cite{DewolfAdaSNNArmCtrl16, ClawsonAdaptiveSNNControl17}, the output connection weights $\mathbf{w}_v$ of the spiking neurons responsible for computing the adaptive linear velocity input are incrementally updated at each time step to minimize the position error $e_p$,
\begin{equation}
    \mathbf{w}_v(t_{k+1})=\mathbf{w}_v(t_k)-\gamma\mathbf{s}_v(t_k)e_p
\end{equation}
where $\gamma$ is the learning rate. Similarly, the angular error signal $e_{\theta}$ is obtained as,
\begin{equation}
    e_{\theta}=\theta-\theta_r
\end{equation}
The output connection weights $\mathbf{w}_w$ responsible for computing the adaptive angular velocity input are incrementally updated to minimize the angular error $e_{\theta}$,
\begin{equation}
    \mathbf{w}_w(t_{k+1})=\mathbf{w}_w(t_k)-\gamma\mathbf{s}_w(t_k)e_{\theta}
\end{equation}
As illustrated in Fig.~\ref{fig:BlockDiagram}, the total control input applied to the vehicle is the sum of the LQR, the feedforward, and the adaptive SNN control signals:
\begin{equation}
    \mathbf{u}(t)=\mathbf{u}_l(t)+\mathbf{u}_f(t)+\mathbf{u}_a(t)
\end{equation}

\section{Results}
\label{sec:Results}
Extensive numerical simulations demonstrate that the adaptive SNN controller outperforms a benchmark LQR controller in wall-following and tracking performance, particularly under modeling uncertainties and external disturbances. This section presents three case studies: straight-line tracking with partial actuator failure (Section~\ref{ssec:ActuatorFail}), sinusoidal trajectory tracking with state estimation errors (Section~\ref{ssec:EstError}), and irregular wall following (Section~\ref{ssec:WF}).

\subsection{Straight-Line Tracking with Actuator Failure}
\label{ssec:ActuatorFail}
\begin{figure}[ht!]
\centering
\includegraphics[width=0.4\textwidth]{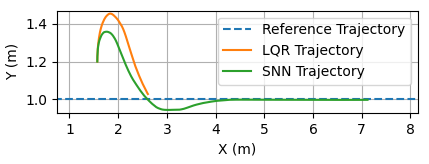}
\caption{Case Study A: straight-line tracking with actuator failure. The adaptive SNN controller converges more quickly to the desired trajectory and covers a greater distance compared to the benchmark LQR controller.}
\label{fig:LineTraj}
\end{figure}

In this case study, partial actuator failure is simulated by reducing the robot's controlled rotational angle by $50\%$. The robot begins at position $(\frac{\pi}{2},1.2)$ with an orientation of $\theta=90\deg$, a constant tracking velocity of $v_r=1$m/s, a maximum angular velocity of $\omega_m=1$rad/s, and a feedforward coefficient of $\alpha=0.06$. If the angular velocity exceeds the threshold, it is limited to the maximum value, and the linear velocity is proportionally reduced. The robot is then instructed to follow the straight line at $y=1$m. Fig.~\ref{fig:LineTraj} compares the trajectories of robots controlled by the LQR and the adaptive SNN controllers over a $20$-second simulation. The adaptive SNN-controlled robot successfully converges to the line after experiencing some overshoot and covers a greater distance, whereas the benchmark LQR controller fails to achieve the control objective within the 20-second timeframe.

\begin{figure}[ht!]
\centering
\includegraphics[width=0.4\textwidth]{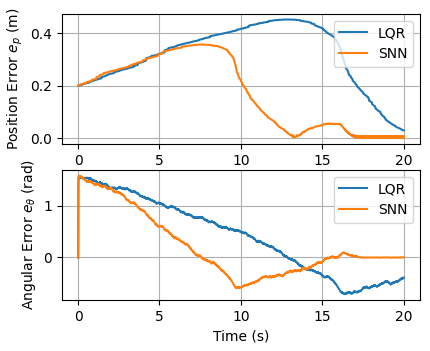}
\caption{Case Study A: the error signals of the adaptive SNN converge to zero at $t=17$s, whereas the LQR fails to achieve convergence by the end of the simulation.}
\label{fig:LineErr}
\end{figure}

\begin{figure}[ht!]
\centering
\includegraphics[width=0.4\textwidth]{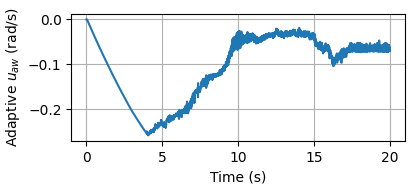}
\caption{Case Study A: the extra control signal for angular velocity generated by the adaptive SNN.}
\label{fig:LineSNNCtrl}
\end{figure}

Fig.~\ref{fig:LineErr} illustrates that the position error ($e_p$) and angular error ($e_{\theta}$) of the adaptive SNN controller converge to zero at approximately $t=17$s, whereas the benchmark LQR controller has not converged by that time. The adaptive SNN controller reaches the reference line more quickly than the LQR controller because it is trained online to adapt to uncertainties and minimize the angular error. As shown in Fig.~\ref{fig:LineSNNCtrl}, the additional control signal for angular velocity generated by the adaptive SNN initially increases when the robot faces the opposite direction and subsequently stabilizes around $-0.07$rad/s to compensate for the angular deviation caused by partial actuator failure. In contrast, the LQR control signal relies entirely on the ideal model and cannot adapt to the imprecise motion resulting from sudden actuator failure. Consequently, the robot controlled by the adaptive SNN tracks the reference straight line significantly faster than the LQR-controlled robot under actuator failure conditions in this case study.

\subsection{Sinusoidal Trajectory Tracking with Estimation Error}
\label{ssec:EstError}
\begin{figure}[ht!]
\centering
\includegraphics[width=0.45\textwidth]{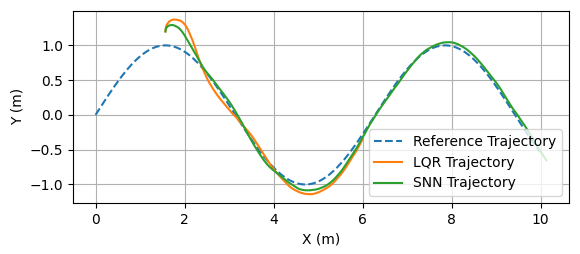}
\caption{Case Study B: sinusoidal trajectory tracking with estimation error. The adaptive SNN controller converges more rapidly to the desired trajectory and covers a greater distance than the benchmark LQR controller.}
\label{fig:SinTraj}
\end{figure}

In this case study, state estimation errors are introduced by  incorporating zero-mean Gaussian noise into the onboard measurements of the robot's position and orientation, with standard deviations of $\sigma_p=0.05$m and $\sigma_{\theta}=0.1$rad, respectively. The robot is configured with the same initial conditions as those described in Case Study A. Subsequently, it is tasked to follow the sinusoidal trajectory defined by $y=\sin x$. Fig.~\ref{fig:SinTraj} compares the trajectories of robots controlled by the LQR and the adaptive SNN controllers over a simulation period of $35$ seconds. Although both controllers began with the robot oriented opposite to the reference matching point, the adaptive SNN controller successfully stabilized the robot and aligned it with the sinusoidal reference trajectory. In contrast, the benchmark LQR-controlled robot required a longer duration to adjust its orientation and exhibited considerably slower movement compared to the adaptive SNN-controlled robot.

\begin{figure}[ht!]
\centering
\includegraphics[width=0.42\textwidth]{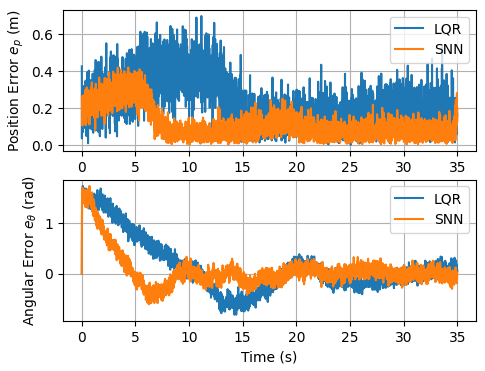}
\caption{Case Study B: the adaptive SNN converges and stabilizes at low constant values within $10$ seconds, whereas the benchmark LQR continues to oscillate substantially.}
\label{fig:SinErr}
\end{figure}

Figure~\ref{fig:SinErr} illustrates that the adaptive SNN controller’s position and angular error signals converge and stabilize at low constant values within $10$ seconds. In contrast, the benchmark LQR’s position error signal continues to oscillate, potentially generating excessive noise and reducing motor lifespan. The adaptive SNN dynamically adjusts the robot’s linear and angular velocities to address state estimation uncertainties and minimize error signals. Conversely, the benchmark LQR assumes perfect state estimation and feedback, making it unable to adapt to uncertainties in real time. Consequently, the adaptive SNN controller tracks the reference sinusoidal trajectory more accurately than the benchmark LQR controller in this case study.

\subsection{Irregular Wall Following}
\label{ssec:WF}
\begin{figure}[H]
    \centering
    \begin{subfigure}[b]{0.23\textwidth}
        \centering
        \includegraphics[width=\textwidth]{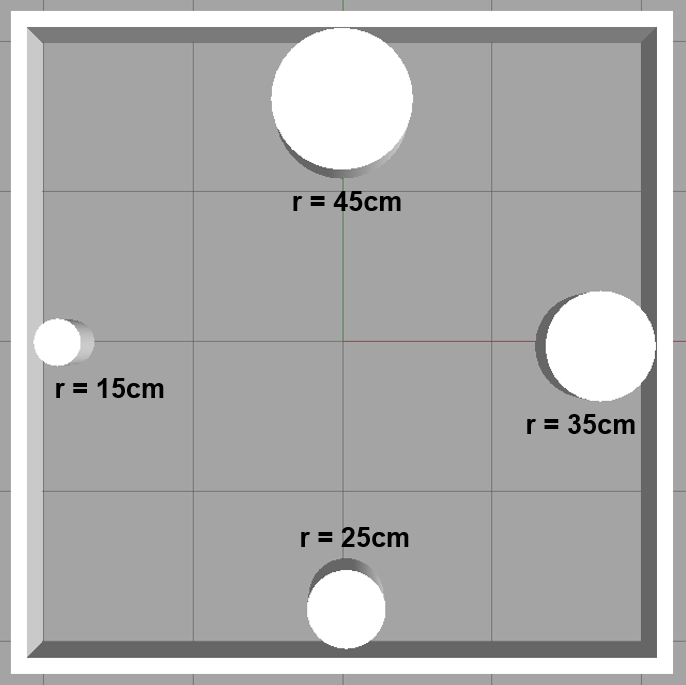}
        \caption{Indoor Environment}
        \label{fig:IndoorEnv}
        \vspace{1mm}
    \end{subfigure}
    \begin{subfigure}[b]{0.23\textwidth}
        \centering
        \includegraphics[width=\textwidth]{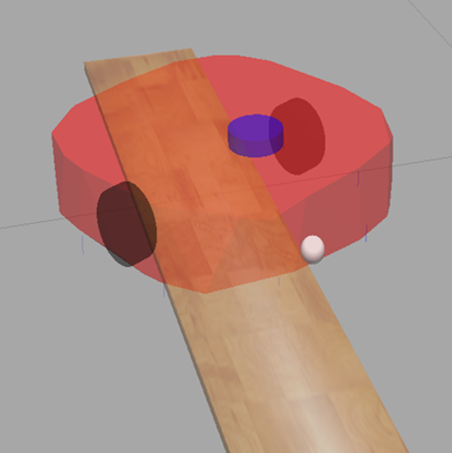}
        \caption{Robot Model}
        \label{fig:EurekaSimModel}
        \vspace{1mm}
    \end{subfigure}
    \begin{subfigure}[b]{0.23\textwidth}
        \centering
        \includegraphics[width=\textwidth]{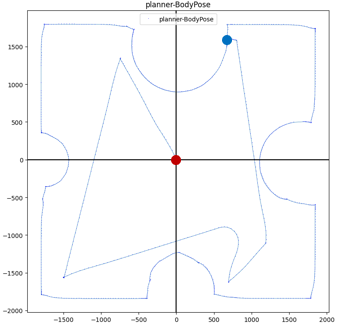}
        \caption{LQR Wall Following}
        \label{fig:WFLQR}
    \end{subfigure}
    \begin{subfigure}[b]{0.23\textwidth}
        \centering
        \includegraphics[width=\textwidth]{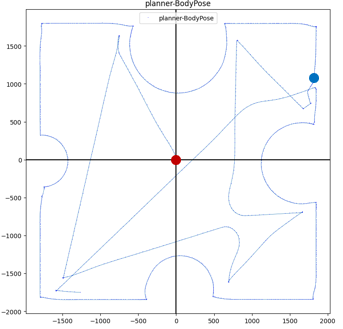}
        \caption{SNN Wall Following}
        \label{fig:WFSNN}
    \end{subfigure}
    \caption{Case Study C: the robot, starting from the center marked by the red dot, explores the room for a designated period before initiating wall-following behavior at the blue dot. The adaptive SNN demonstrates superior accuracy in tracking circular walls compared to the benchmark LQR.}
    \label{fig:WFTraj}
\end{figure}

As shown in Fig.~\ref{fig:EurekaSimModel}, a LiDAR-equipped two-wheeled robotic vacuum cleaner~\cite{EurekaIntro} is simulated with high fidelity in Gazebo to navigate and clean a $4\times4$ meters square area containing four cylinders of varying radii positioned along the walls. The robot is programmed to start at the center of the room, initially explore the unknown space, and subsequently engage in wall-following behavior while maintaining a distance of $d=180$mm from the walls. Fig.~\ref{fig:WFLQR} and \ref{fig:WFSNN} compares the performance of the benchmark LQR controller with that of the adaptive SNN controller. The results demonstrate that the adaptive SNN tracks the circular contours of the walls more accurately than the benchmark LQR controller, achieving a mean absolute error of $0.06$ meters compared to $0.11$ meters for the LQR. This enhanced accuracy in contour tracking suggests that the adaptive SNN can significantly improve the operational efficiency of robotic vacuum cleaners in complex environments.

\section{Conclusions}
\label{sec:Conclusion}
This paper presents an adaptive SNN-based wall-following control framework that learns and accounts for modeling uncertainties in real-time. Initially, B-spline fitting is utilized to model the unknown wall shape and generate the reference trajectory for the LQR to track. An optimal discretized LQR control solution is derived based on an ideal vehicle model. Additionally, we develop point-matching algorithms to produce a feedforward control signal that accelerates convergence. Subsequently, an SNN-based feedback controller is designed to incrementally adjust its connection weights in response to error signals, thereby adapting to significant uncertainties. Through extensive numerical simulations, our adaptive SNN control design demonstrates superior performance over the benchmark LQR in terms of tracking accuracy and convergence speed, especially under uncertainties such as partial actuator failures and state estimation errors.

\addtolength{\textheight}{-12cm}  % This command serves to balance the column lengths

\bibliographystyle{ieeetr}
\bibliography{adaptive_snn}

\end{document}